\documentclass[twocolumn,showkeys,prd,nofootinbib,floatfix,preprintnumbers]{revtex4-1}

\usepackage[utf8]{inputenc}
\usepackage{multirow}
\usepackage{amsfonts,amsmath,amssymb} 
\usepackage{graphicx,graphics,color}
\usepackage{gensymb}
\usepackage{longtable}
\usepackage{subfigure}
\usepackage{amsmath}

\usepackage{hyperref}
\usepackage[normalem]{ulem}
\hypersetup{
    colorlinks=true,
    linkcolor=blue}

\newcommand{\be}{\begin{equation}}
\newcommand{\ee}{\end{equation}}
\newcommand{\ba}{\begin{eqnarray}}
\newcommand{\ea}{\end{eqnarray}}

\begin{document}

\title{Perturbative Effects of Dark Matter Environments on Black Hole Shadows}

%\title{Perturbing the  Black Hole Shadows with generic dark matter profiles}

\author{Gabriel G\'omez}
\email{luis.gomezd@umayor.cl}
\affiliation{Centro Multidisciplinario de F\'isica, Vicerrector\'ia de Investigaci\'on, Universidad Mayor, \\ Camino La Pir\'amide 5750,  Huechuraba, 8580745, Santiago, Chile}

\begin{abstract}

Constructing spacetime solutions that describe black holes embedded in dark matter environments is a crucial step toward probing the properties of dark matter in the strong–field regime of gravity. At present, however, there is no unique or systematic framework to model such configurations, and several commonly adopted approaches raise methodological ambiguities. 
Motivated by these challenges, we build upon a perturbative framework to describe deformations of static, spherically symmetric black holes induced by a surrounding dark matter distribution. Within this framework, we compute the leading-order corrections to both the photon-sphere radius and the radius of the black hole shadow, assuming a generic dark matter halo profile. We then apply the formalism to physically motivated density profiles, including the Hernquist and Navarro–Frenk–White models, obtaining closed-form analytical expressions for the perturbed metric functions and for the critical impact parameter in the Schwarzschild background. Using these results, we obtain quantitative estimates for the corresponding shadow deviations and find that they lie well beyond the current observational bounds set by Keck and VLTI measurements. As a consistency check, we further estimate the total dark matter mass enclosed within the orbital radius of the S2 star and show that it remains well below the $0.1\%$ upper limit reported by the GRAVITY collaboration. Overall, this approach offers a systematic avenue to investigate perturbative effects of dark matter on black hole phenomenology, including potential implications for gravitational wave observations.
%it is important to check this formalism as a long-program in black hole physics. future high-precision measurements will demand a more refine modeling of the dark matter spike.
%Although the resulting deviations are well below current observational sensitivities, our analysis provides a consistent theoretical baseline for future high-precision shadow measurements. Such data may require a refined treatment, including higher-order perturbative corrections and more general matter configurations.

\end{abstract}

\maketitle

\section{Introduction}

%%%%%%%%%%%%%%%%%%%%%%%%%%%%%%%%%%%%%%%%%%%%%%%%%%%%%%%%%%%%%%%%
%%%%%%%%%%%%%%%%%%%%%%%%%%%%%%%%%%%%%%%%%%%%%%%%%%%%%%%%%%%%%%%%

Current observational efforts such as the Event Horizon Telescope (EHT)
\cite{EventHorizonTelescope:2019dse,2022ApJ...930L..17E,EventHorizonTelescope:2022wkp}, the GRAVITY collaboration \cite{GRAVITY:2020gka}, and the LIGO–VIRGO
collaborations \cite{Abbott:2016blz,LIGOScientific:2017ync} have provided
compelling evidence for the existence of black holes (BHs) and opened a
promising avenue to test fundamental predictions of gravity in the strong–field regime.

Black holes constitute ideal laboratories to study high–energy astrophysical
phenomena occurring in extreme gravitational environments. Moreover, they
encode information about the surrounding matter distribution through the
structure of the spacetime geometry. In this context, BH observations offer a
unique opportunity to probe the properties of the still elusive dark matter
(DM), for instance through possible deviations from the standard vacuum
Schwarzschild and Kerr paradigms.

Of particular relevance within this program is the observation of black hole shadows, whose size and shape are primarily determined by the black hole mass and spin within general relativity (GR). For a Schwarzschild black hole, the shadow is perfectly circular with a radius fixed by the mass, while in the Kerr case it remains nearly circular for moderate spins but develops characteristic distortions as the spin increases \cite{Perlick:2021aok}. Therefore, any measurable deviation from the GR prediction—once mass and spin are independently constrained—may signal additional gravitational effects, potentially arising from a surrounding dark matter distribution. In this context, dark matter halos could leave observable imprints on shadow observables (see e.g. Refs.~\cite{Konoplya:2019sns,Cardoso:2021wlq,Xavier:2023exm,Macedo:2024qky,Gomez:2024ack,Yang:2023tip,Ma:2022jsy,Capozziello:2023tbo,Vertogradov:2024qpf,Pantig:2024rmr,Donmez:2024luc,Ovgun:2025bol,Lobo:2025kzb,Fonseca:2025ehf,Saurabh:2020zqg,Anjum:2023axh}). Such effects are particularly relevant in light of the ring-like images of M87$^\star$ and Sgr A$^\star$ obtained by the EHT \cite{2022ApJ...930L..17E}, which provide direct access to horizon-scale observables.

Remarkably, current observations are fully consistent with GR predictions
based on vacuum Schwarzschild and Kerr geometries (see e.g.
\cite{Vagnozzi:2022moj}). This implies that any deviation induced by a surrounding dark matter distribution must lie within present observational uncertainties. In other words, the impact of DM on black hole phenomenology is expected to be small. Nevertheless, given the current error bars and with the prospect of future high–precision measurements of black hole shadows with the next-generation EHT (ngEHT) program \cite{Johnson:2023ynn,Doeleman:2023kzg}, there remains significant potential to probe dark matter effects in the strong–field regime.

These considerations naturally motivate the use of perturbative approaches to
estimate deviations from vacuum black hole geometries. Such methods allow one
to capture the relevant physical effects without resorting to fully numerical
and computationally expensive modeling of the complete BH–DM system.
Perturbative schemes can therefore provide both reliable analytical insight
and valuable inputs for future high–precision numerical studies of the full
spacetime geometry \cite{Babichev:2012sg,Kimura:2021dsa,LaHaye:2025ley,Gliorio:2025cbh,Dyson:2025dlj}.

A perturbative framework to describe deformations of static, spherically symmetric black holes induced by surrounding matter distributions was introduced in \cite{Gomez:2024ack}. In that work, the formalism was developed and applied to self--interacting scalar--field dark matter around a Schwarzschild black hole. In the present work, we substantially extend this framework in three directions. First, we demonstrate that the formalism is applicable to generic dark matter environments and, more generally, to arbitrary static, spherically symmetric black hole spacetimes, leading to a general analytical expression for the critical impact parameter beyond the Schwarzschild solution. Second, we clarify the physical conditions governing the validity of the perturbative expansion by showing that it is controlled by the smallness of the induced metric deformations, rather than by the dark matter mass itself. Third, we assess the consistency and range of applicability of the perturbative framework by evaluating the metric deformations at all physically relevant characteristic radii. These developments enable a systematic and analytically tractable treatment of realistic phenomenological density profiles, such as the Hernquist and Navarro--Frenk--White distributions, allowing us to quantify strong--field environmental effects on black hole shadows and related observables.

Most existing studies modeling the spacetime of a black hole embedded
in a dark matter halo have relied on Newtonian–based treatments, in
which the metric coefficients are reconstructed from the tangential
velocity profile together with an assumed dark matter density
distribution \cite{Matos:2003nb,Xu:2018wow}. Other works
\cite{Cardoso:2021wlq,Konoplya:2022hbl,Macedo:2024qky} have instead
investigated solutions of the Einstein equations sourced by anisotropic
halos, typically characterized by purely tangential pressure.

However, growing concerns have been raised regarding the underlying
assumptions adopted in these approaches, which point to the absence of
a systematic and fully self–consistent framework for describing
black hole–dark matter spacetimes. We refer the reader to
Refs.~\cite{Datta:2023zmd,Zhao:2026yis,Fauzi:2025yse,Bolokhov:2025zva}
for detailed discussions of these issues.

In this work, we do not aim to revisit or assess these existing approaches. Instead, we adopt an alternative strategy based on the standard perturbation theory. This framework provides a systematic and physically transparent way to construct BH–DM spacetimes, ensuring consistency with GR while retaining analytical tractability.

%For instance, in the Einstein cluster method, where the environment is modeled through an anisotropic stress–energy tensor, it is commonly assumed that the radial pressure vanishes. However, recent studies have pointed out that this assumption may not be physically justified in realistic astrophysical environments and could lead to inaccuracies when modeling environmental effects on gravitational wave signals \cite{Datta:2023zmd,Zhao:2026yis}.  Alternative constructions have employed effective equations of state with negative pressure to describe dark matter halos. Such treatments, however, can lead to qualitatively different spacetime structures, including contrasting predictions for photon sphere properties and shadow observables \cite{Fauzi:2025yse}. More recently, it has been emphasized that some proposed BH–DM metrics rely on assumptions that may violate standard energy conditions or fail to satisfy the Einstein equations in a self–consistent manner, thereby casting doubt on their reliability as physically viable solutions \cite{Bolokhov:2025zva}.

The paper is organized as follows. In Sec.~\ref{sec:Perturbative}, we revisit the perturbative framework and describe the properties of metric deformations induced by dark matter distributions, as well as the corresponding corrections to the photon sphere and the black hole shadow radius. In Sec.~\ref{sec:applications}, we apply the formalism to specific dark matter density profiles and discuss the resulting constraints and physical implications. Finally, in Sec.~\ref{sec:Conclusions}, we summarize our findings and outline possible directions for future work.

\section{Perturbative framework for black holes in dark matter environments}
\label{sec:Perturbative}

In this section, we review the perturbative framework introduced in Ref.~\cite{Gomez:2024ack}, which provides a systematic method to describe metric deformations induced by dark matter distributions surrounding a black hole. Building upon this framework, we derive general analytical expressions for the leading-order corrections to the photon-sphere radius and to the black hole shadow. These results ultimately allows us to derive constraints on
dark matter properties from EHT shadow–size measurements.

\subsection{Perturbed Einstein equations}

%We expect any deviation from the Schwarzschild BH to be very small, including the one due to DM environment. Then we need to solve the Einstein equation in the form 
%\begin{equation}
 %   G^\mu{}_\nu = 8\pi{\cal G} \epsilon %T^\mu{}_\nu,
%\end{equation}
%with the small parameter $\epsilon$. At zeroth order, the metic is given by a vacuum solution of the Einstein equations. Once one considers the effect of the environment encoded in $ \epsilon T^\mu{}_\nu$, the spacetime will be described by the metric 
%\begin{equation}
 %   g_{\mu\nu} =g_{\mu\nu}^{\rm %Sch}+\epsilon \delta g_{\mu\nu},
%\end{equation}
We consider a static, spherically symmetric spacetime described by
\begin{equation}
ds^2 = -f(r)\,dt^2 + g(r)\,dr^2 + r^2 d\Omega^2 ,\label{eqn:metric}
\end{equation}
where $d\Omega^2$ is the metric on the unit two--sphere. 
Since any deviation from a vacuum black hole is expected to be small, including those induced by a surrounding dark matter environment, we adopt a perturbative treatment of the Einstein equations. We expand the metric around a Schwarzschild background,
\begin{align}
f(r) &= f_0(r) + \delta f(r), \nonumber \\
g(r) &= g_0(r) + \delta g(r),\label{eqn:expansion}
\end{align}
with
\begin{equation}
f_0(r) = 1 - \frac{2{\cal G}M_0}{r},
\qquad
g_0(r) = \frac{1}{f_0(r)} .
\label{eqn:metric_back}
\end{equation}
The perturbations $\delta f(r)$ and $\delta g(r)$ encode the gravitational backreaction of the dark matter distribution onto the spacetime metric. Here $M_{0}$ is the black hole mass. 

It is convenient to introduce the Misner--Sharp mass function,
\begin{equation}
m(r) = M_0 + \delta m(r),
\end{equation}
where $\delta m(r)$ represents the contribution of the dark matter distribution. At linear order, this uniquely determines the perturbation of the radial metric
components,
\begin{equation}
\delta g(r) = \frac{2{\cal G}\,\delta m(r)}{r\,f_0(r)^2}.\label{eq:delta_g}
\end{equation}
This is therefore a perturbative identity arising from the expansion around a vacuum background rather than a property of the full non–vacuum spacetime.

It is important to emphasize that the perturbative expansion is controlled by the smallness of the metric deformations induced by the dark matter distribution,
\begin{equation}
\frac{\delta f(r)}{f_{0}(r)}\ll1,
\qquad
\frac{\delta g(r)}{g_0(r)}
=
\frac{2{\cal G}\,\delta m(r)}{r\,f_0(r)}
\ll1.\label{eqn:conditions}
\end{equation}
rather than by the dark matter mass itself.
These conditions ensure that the spacetime remains a small deformation of the Schwarzschild geometry in the region of interest. Hence, higher--order nonlinear effects can be safely neglected. Consequently, the perturbative treatment remains valid even when $\delta m(r)$ becomes comparable to or exceeds the black hole mass, provided the dark matter distribution is sufficiently extended so that the induced metric deformation remains small. This situation is naturally realized for realistic galactic dark matter halos,
such as those described by Hernquist or NFW--type profiles. An alternative possibility, not explored here, is a dark matter configuration
that is both sufficiently light and compact, for which the same perturbative conditions would also be satisfied.

On the other hand, the metric perturbations given by Eqs.~(\ref{eqn:expansion}) are not valid arbitrarily close to the event horizon, where the background metric function $f_{0}(r)$ vanishes. For this reason, our analysis is restricted to radii greater than or equal to the photon radius, ensuring both physical relevance and perturbative control.

The linearized Einstein equations $\delta G^\mu{}_\nu = 8\pi{\cal G} T^\mu{}_\nu$
yield the following independent relations:
\begin{align}
\frac{\partial \delta m}{\partial r} &= -4\pi r^2 T^{t}{}_{t},
\label{eqn:tt}\\
\frac{\partial \delta m}{\partial t} &= 4\pi r^2 T^{r}{}_{t},
\label{eqn:rt}\\
- \frac{2 {\cal G} \delta m}{r^2 f_0}
+ \frac{\partial \delta f}{\partial r}
- \frac{2 {\cal G} M_0}{r^2 f_0}\,\delta f
&= 8 \pi {\cal G} r\, T^{r}{}_{r},
\label{eqn:rr}
\end{align}
where $T^\mu{}_\nu$ denotes the dark matter energy–momentum tensor that sources the metric perturbations.

In this work we assume for simplicity that the dark matter halo is static on the timescales of interest and does not accrete onto the black hole. This implies a vanishing
radial energy flux,
\begin{equation}
T^{r}{}_{t} = 0,
\end{equation}
so that Eq.~\eqref{eqn:rt} enforces $\partial_t \delta m = 0$. The perturbative
mass function is therefore purely radial. We further take the stress--energy tensor in the diagonal form
\begin{equation}
T^{\mu}{}_{\nu} = \mathrm{diag}(-\rho,\,P_r,\,0,\,0).
\end{equation}
This form is adopted for consistency with the symmetry of the perturbative
setup. Since the metric perturbations considered here are purely radial and the linearized Einstein equations involve only the $tt$ and $rr$ components, tangential stresses do not enter at this order. 

Equation~\eqref{eqn:tt} then reduces to
\begin{equation}
\frac{d\,\delta m}{dr} = 4\pi r^2 \rho(r),
\label{eqn:deltam}
\end{equation}
which determines $\delta m(r)$ directly from the density profile.

Combining the linearized $tt$ and $rr$ equations, and assuming stationarity of
the metric perturbations, $\partial_t \delta f = 0$, Eq.~\eqref{eqn:rr} can be
cast into the first–order radial equation
\begin{equation}
\frac{d}{dr}\!\left(\frac{\delta f}{f_0}\right)
=
\frac{2{\cal G}}{r^2 f_0^2}
\left[
\delta m(r) + 4\pi r^3 P_r(r)
\right].
\label{eq:dfmaster}
\end{equation}

\subsection{Perturbed metric functions}

Integrating Eq.~\eqref{eqn:deltam} and fixing the integration constant by imposing that the perturbative mass approaches a constant outside the halo radius  $R_{\rm halo}$, namely
\begin{equation}
r \geq R_{\rm halo}:\qquad
\delta m(r) = \delta M_0 ,
\end{equation}
with
\begin{equation}
    \delta M_0\equiv 4\pi\!\int_{r_{\min}}^{R_{\rm halo}} dr'\, r'^2 \rho(r'),
\end{equation}
one obtains the general solution
\begin{equation}
\delta m(r)
=
\delta M_0
-
4\pi\!\int_r^{R_{\rm halo}} dr'\, r'^2 \rho(r').
\end{equation}
This result admits a transparent physical interpretation: the dark matter mass enclosed within radius $r$ is given by the total mass of the distribution minus the mass located outside that radius. Equivalently, Eq.~\eqref{eqn:deltam_sol}
can be written in the standard enclosed--mass form
\begin{equation}
\delta m(r) = 4\pi\!\int_{r_{\min}}^{r} dr'\, r'^2 \rho(r').\label{eqn:deltam_sol}
\end{equation}
The lower cutoff $r_{\min}$ is chosen slightly outside the horizon, where the perturbative expansion remains well defined. The condition $\delta m(r_{\min})=0$, setting by construction, ensures that the central black hole mass remains unchanged, while $\delta m(r\ge R_{\rm cloud})=\delta M_0$ guarantees a smooth matching to an exterior vacuum solution.

Integrating Eq.~\eqref{eq:dfmaster} and imposing asymptotic flatness,
\begin{equation}
\delta f(r\to\infty)=0,
\end{equation}
one finds
\begin{equation}
\delta f(r)
=
-2{\cal G}\, f_0(r)
\int_r^\infty
\frac{dr'}{r'^2 f_0(r')^2}
\left[
\delta m(r') + 4\pi r'^3 P_r(r')
\right].
\label{eqn:deltaf_sol}
\end{equation}
The second term inside the brackets represents a relativistic correction due to radial pressure. In this work, we neglect this contribution, as we remain agnostic about the microscopic nature of dark matter and focus on phenomenological halo models.  As a consistency check, one has
\begin{equation}
r \geq R_{\rm halo}:
\delta f(r) = -\frac{2{\cal G}\delta M_0}{r},\quad \delta g(r) = \frac{2{\cal G}\delta M_0}{r f_{0}^2}, \label{eq:metric_edge}
\end{equation}
demonstrating that the spacetime exterior solution is exactly Schwarzschild with a
shifted mass,
\begin{equation}
M_{\rm tot} = M_0 + \delta M_0.
\end{equation}

\subsection{Black-hole shadow in dark-matter environments}

A key observable associated with black holes is the so–called
\emph{shadow}, namely the dark region in the observer’s sky
corresponding to photon trajectories that are strongly bent by the
black hole’s gravitational field. Its boundary is determined by
unstable circular null geodesics, which define the photon sphere, sometimes referred to as light rings. For a detailed discussion and derivation, see Ref.~\cite{Perlick:2021aok}.
The properties of these orbits encode direct information about the spacetime geometry in the strong–field regime
\cite{EventHorizonTelescope:2020qrl,Volkel:2020xlc}.

For a static, spherically symmetric metric of the form given by Eq.~(\ref{eqn:metric}), the photon sphere radius $r_{\rm ph}$ is obtained from the condition of the effective potential for null geodesics \cite{Bardeen:1973tla}, which yields
\begin{equation}
r_{\rm ph} =
2 f(r) \left(\frac{df}{dr}\right)^{-1}
\bigg|_{r=r_{\rm ph}}.
\label{eqn:r_photon}
\end{equation}

The observable size of the shadow is determined by the critical
impact parameter $b_{\rm cr}$ associated with unstable photon
orbits. From the perspective of a distant observer, light rays can be classified into two distinct families: photons with $b < b_{\rm cr}$
are captured by the black hole, while those with $b > b_{\rm cr}$
are deflected by the gravitational field and eventually escape to infinity. The critical
impact parameter can be written as \cite{Psaltis:2007rv}
\begin{equation}
b_{\rm cr} =
\frac{r_{\rm ph}}{\sqrt{f(r_{\rm ph})}} .
\label{bcr}
\end{equation}
For a distant observer located at $r \gg R_{\rm halo}$, the angular
shadow radius is simply
\begin{equation}
\alpha_{\rm sh} = \frac{b_{\rm cr}}{r} .
\end{equation}
Hence, the problem reduces to determining how the photon sphere and
the corresponding critical impact parameter are modified by the dark
matter environment.

\medskip

At linear order in the metric perturbation, the shift of the photon sphere
follows from Eq.~(\ref{eqn:r_photon}) by expanding both the metric function
and the photon-sphere radius around their background values. One finds
\begin{equation}
\delta r_{\rm ph}
=
\frac{
2 \delta f(r_{\rm ph0})
-
r_{\rm ph0} \delta f'(r_{\rm ph0})
}{
r_{\rm ph0} f_{0}''(r_{\rm ph0})
-
f_{0}'(r_{\rm ph0})
},
\end{equation}
where a prime denotes differentiation with respect to $r$. At this stage, this expression is completely general and applies to any
static, spherically symmetric geometry.

At zeroth order, the background solution is Schwarzschild, for which
\begin{equation}
r_{\rm ph0}^{\rm Sch} = 3 {\cal G} M_0 = \frac{3}{2} r_s ,
\qquad
r_s = 2 {\cal G} M_0 ,
\end{equation}
with $r_s$ the Schwarzschild radius of the black hole. 

To make the physical content of the correction more transparent,
we now use the linearized Einstein equations, which allow us to express
the corrected photon-sphere radius as
\begin{equation}
r_{\rm ph}^{\rm Sch}
=
r_{\rm ph0}^{\rm Sch}
\left[
1
+
\frac{\delta m(r_{\rm ph0})}{M_0}
+
4\pi {\cal G} r_{\rm ph0}^2
P(r_{\rm ph0})
\right].
\label{eqn:photon_sphere}
\end{equation}

This expression admits a clear physical interpretation. The term
$\delta m(r_{\rm ph0})$ represents the additional mass enclosed
within the photon–sphere radius due to the dark matter distribution.
The pressure term $P(r_{\rm ph0})$ constitutes a genuinely
relativistic contribution that vanishes for pressureless dust but becomes relevant for more general equations of state.

%Importantly, Eq.~(\ref{eqn:deltam_sol}) implies that, provided $r_s < r_{\min} < r_{\rm ph0}$, the photon sphere location depends only on the mass distribution interior to $r_{\rm ph}$. Consequently, the detailed structure of the halo at larger radii does not influence this strong–field observable.

\medskip

The corresponding critical impact parameter, expanded to first order,
takes the general form
\begin{equation}
b_{\rm cr}
=
b_{\rm cr0}
\Bigg[
1
+
\left(
\frac{2f_{0}(r_{\rm ph0}) - r_{\rm ph0} f_{0}'(r_{\rm ph0})}
{2 r_{\rm ph0} f_{0}(r_{\rm ph0})}
\right)
\delta r_{\rm ph}
-
\frac{\delta f(r_{\rm ph0})}{2 f_{0}(r_{\rm ph0})}
\Bigg],
\end{equation}
which is valid for any static, spherically symmetric background geometry and generic DM distribution.

We now specialize to the Schwarzschild solution, for which
\begin{equation}
b_{\rm cr0}^{\rm Sch} = 3\sqrt{3}\,{\cal G}M_{0},
\end{equation}
and the photon-sphere radius satisfies the identity $ 
2 f_{0}(r_{\rm ph0}) = r_{\rm ph0} f_{0}'(r_{\rm ph0})$. As a consequence, the term proportional to $\delta r_{\rm ph}$ vanishes,
and the corrected critical impact parameter simplifies to
\begin{equation}
b_{\rm cr}^{\rm Sch}
=
b_{\rm cr0}^{\rm Sch}
\left[
1
-
\frac{3}{2}\,\delta f(r_{\rm ph0})
\right].\label{eqn:bcr_pert}
\end{equation}
Hence, deviations of the shadow size from the Schwarzschild prediction are controlled entirely by the metric perturbation evaluated at the photon sphere. This highlights that shadow observables probe the dark-matter distribution in the immediate strong–field region, rather than the global halo structure.

\subsection{EHT Shadow-size measurements}

Psaltis \textit{et al.}~\cite{EventHorizonTelescope:2020qrl} proposed using the observed shadow size of M87$^\star$ as a probe of possible deviations from the predictions of GR. Current EHT measurements indicate that the
shadow diameter is consistent, within $\sim 17\%$ at the $68\%$ confidence level, with the expectation for a Schwarzschild black hole. The same
observational benchmark can therefore be employed to constrain additional
gravitational effects arising from a surrounding dark matter environment, whose presence could induce small but potentially measurable departures from
the vacuum Schwarzschild prediction.

The resulting fractional deviation of the shadow radius\footnote{In the remainder of the paper, we drop the superscript on all quantities, as it is understood that we focus on the Schwarzschild case.}, $
\delta \equiv (b_{\rm cr}^{\rm Sch}-b_{\rm cr0}^{\rm Sch})/b_{\rm cr0}^{\rm Sch}$, can be directly confronted with current observational constraints. In particular, measurements from the
Keck Observatory and the VLTI yield \cite{EventHorizonTelescope:2022xqj}
\begin{equation}
\text{Keck:}\quad
\delta =- 0.04^{+0.09}_{-0.10},
\end{equation}
\begin{equation}
\text{VLTI:}\quad
\delta = -0.08^{+0.09}_{-0.09},
\end{equation}
which translate into bounds on the allowed magnitude of
$\delta f(r_{\rm ph0})$, thereby probing dark matter properties at the characteristic scale of the photon sphere.

\section{Applications to realistic dark-matter halo models}\label{sec:applications}

For illustrative purposes, we consider the NFW and Hernquist profiles, both of
which provide accurate descriptions of dark matter haloes in cosmological
simulations. The two profiles share a common inner behavior, exhibiting a cusp
that scales as $\rho\propto r^{-1}$. They differ, however, in their outer
structure: the Hernquist profile falls off more steeply,
$\rho\propto r^{-4}$, whereas the NFW profile decays as $\rho\propto r^{-3}$.
Consequently, unlike the NFW profile, the Hernquist profile has a finite total
mass, allowing one to model isolated haloes without introducing an ad hoc
truncation (see e.g., \cite{Springel:2004kf}).

In this work, we do not impose a specific halo mass definition such as
$M_{200}$ within $r_{200}$, commonly used in numerical simulations and defined as the mass enclosed at which the mean density equals 200 times the critical density of the Universe.
Instead, we deliberately introduce an arbitrary halo radius $R_{\rm halo}$ to
parameterize the cumulative enclosed mass. While this choice introduces a mild
ambiguity in the global mass normalization, it does not affect our results. Our analysis focuses
on the strong--field region near the photon sphere, where the metric perturbations
are controlled solely by the local enclosed mass at small radii and are
insensitive to the precise definition of the halo boundary. In this sense,
$R_{\rm halo}$ should be regarded as a bookkeeping scale rather than a physical
cutoff, and any reasonable choice leads to identical conclusions in the regime
of interest.

\medskip
Although the dark matter distribution is expected to be modified in the immediate vicinity of the black hole—most notably through the formation of a density spike driven by adiabatic accretion—our goal here is not to model these relativistic mechanisms in detail. Rather, our purpose is to assess the internal consistency and domain of validity of the perturbative scheme itself.

Remarkably, within this approach the resulting mass distribution remains finite for radii $r\gtrsim r_{\rm ph0}$, without the need to introduce ad hoc inner truncation factors that are often employed in black holes-dark matter systems (see, e.g., Ref.~\cite{Cardoso:2021wlq}). This allows us to consistently probe the impact of extended dark matter environments on strong-field observables while remaining agnostic about the detailed relativistic effects governing the innermost region, which has been studied elsewhere \cite{Gondolo:1999ef,Sadeghian:2013laa,Kamermans:2024ieb,Chakraborty:2024gcr}.

\subsection{Hernquist density profile}

We consider a dark matter halo described by a Hernquist density profile \cite{Hernquist:1990be}
\begin{equation}
\rho(r)=\frac{M_{\rm halo}}{2\pi}\,
\frac{a}{r(r+a)^3},
\label{eq:Hernquist_rho}
\end{equation}
where $M_{\rm halo}$ is the total halo mass and $a$ is the scale radius, determining the transition from its inner slope to its outer slope.

Integrating the density profile and imposing $\delta m(r_{\rm ph0})=0$, one finds the enclosed perturbative mass
\begin{equation}
\delta m(r)
=
M_{\rm halo}
\left[
\frac{r^2}{(r+a)^2}
-
\frac{r_{\rm ph0}^2}{(r_{\rm ph0}+a)^2}
\right].
\label{eq:deltam_hernquist}
\end{equation}
The subtraction term reflects the exclusion of the region $r\leq r_{\rm ph0}$,
ensuring that the perturbation vanishes at the photon sphere. Notice that in the asymptotic limit $r\to\infty$ the enclosed mass satisfies
$\delta m(r)\to M_{\rm halo}$. However, at any finite radius—including the halo boundary—the enclosed mass differs from the total halo mass,
$\delta m(r)\neq M_{\rm halo}$, as explicitly illustrated below.

The integral for $\delta f$, Eq.~(\ref{eqn:deltaf_sol}), can be evaluated in closed form,
yielding
\begin{align}
\delta f(r)
=&-2\,{\cal G}\,M_{\rm halo}\,f_0(r)
\Bigg[
-\frac{r_{\rm ph0}^2}{(a+r_{\rm ph0})^2\,(r-r_s)}
\nonumber\\[4pt]
&\quad
+\frac{1}{(a+r_s)^3}
\Bigg(
\frac{(a+r_s)\!\left[a^2(r- r_s)+(a+r)r_s^2\right]}
{(a+r)(r-r_s)}
\nonumber\\[4pt]
&\qquad\quad
+\,2a r_s
\ln\!\left(\frac{a+r}{r-r_s}\right)
\Bigg)
\Bigg].
\label{eq:deltaf_hernquist}
\end{align}
For completeness, we also provide the expression for $\delta g(r)$ (see Eq.~(\ref{eq:delta_g}), which
follows directly from the perturbative mass function in
Eq.~(\ref{eq:deltam_hernquist}). It is given by
\begin{equation}
\delta g(r)
= \frac{2\,{\cal G}\;M_{\rm halo}}{r f_{0}^{2}}
\left[
\frac{r^2}{(r+a)^2}
-
\frac{r_{\rm ph0}^2}{(r_{\rm ph0}+a)^2}
\right].
\label{eq:deltag_hernquist}
\end{equation}
These expressions fully characterize the
first-order deformation of the Schwarzschild geometry induced by a Hernquist dark matter halo.

Before turning to the fractional deviation of the shadow, we first examine representative regions of interest corresponding to given hierarchical scales, which provide a direct consistency check of the perturbative expansion.

\subsubsection{Halo-edge limit} 
At the halo boundary, $r=R_{\rm halo}$, and under the hierarchy $r_{\rm ph0}\ll a,R_{\rm halo}$, while keeping the ratio $R_{\rm halo}/a$ finite, one finds
\begin{equation}
\begin{aligned}
\delta M_0
=  \delta m(R_{\rm halo}) = & M_{\rm halo}\left(\frac{R_{\rm halo}^{2}}{(a+ R_{\rm halo})^{2}} - \frac{r_{\rm ph0}^{2}}{a^{2}}\right) & + \\ \mathcal{O}\left(\frac{r_{\rm ph0}^{3}}{a^{3}}\right).
\end{aligned}
\end{equation}
This result shows that the mass contribution relevant for the perturbative expansion is controlled by the fraction of the halo enclosed between
$r_{\rm ph0}$ and $R_{\rm halo}$, as expected. Moreover, the correction term is parametrically suppressed by
$r_{\rm ph0}/a\ll1$.

Although realistic Hernquist halos typically satisfy
$R_{\rm halo}\sim a\sim \mathcal{O}(\rm kpc)$, it is nevertheless instructive to consider two limiting cases
as consistency checks.

In the regime $R_{\rm halo}\ll a$, and keeping the leading
correction induced by the finite inner cutoff, the total perturbative mass
inside the cloud simply reduces to
\begin{equation}
\delta m(R_{\rm halo})
\approx
M_{\rm halo}
\left(
\frac{R_{\rm halo}^2}{a^2}
-
\frac{r_{\rm ph0}^2}{a^2}
\right) + \mathcal{O}\left(\frac{r_{\rm ph0}^{3}}{a^{3}}\right).
\end{equation}
Thus, the enclosed mass scales as $\delta M_0\propto R_{\rm halo}^2$, as expected from the inner $1/r$ cusp of the Hernquist density profile.

In the opposite limit, $R_{\rm halo}\gg a$, one finds
\begin{equation}
\begin{aligned}
\delta m(R_{\rm halo})
\approx &
M_{\rm halo}
\left(
1-\frac{2a}{R_{\rm halo}}+\frac{3a^{2}}{R_{\rm halo}^{2}}-\frac{r_{\rm ph0}^2}{a^2}
\right) +  \\ & \mathcal{O}\left(\frac{r_{\rm ph0}^{3}}{a^{3}},\frac{a^{3}}{R_{\rm halo}^{3}}\right).
\end{aligned}
\end{equation}
Although in this regime $\delta M_0$ approaches the total halo mass, this does not by itself invalidate the perturbative expansion. We emphasize that the validity of the scheme is controlled by the conditions set in Eq.~(\ref{eqn:conditions}), rather than by the ratio $\delta M_0/M_0$, as one might naively expect. For an extended halo with $a\gg r_{\rm ph0}$, the relevant compactness
${\cal G}\,\delta m(r)/r$ remains small at all radii of interest, ensuring that the spacetime is a mild deformation of the Schwarzschild geometry even when
$M_{\rm halo}\gg M_0$.

Similarly, at the halo edge, $r=R_{\rm halo}$, with $r_{\rm ph0}\ll R_{\rm halo},a$, the leading contribution to the
metric perturbations read
\begin{equation}
  \begin{aligned}
\delta f(R_{\rm halo})
\simeq & -\frac{2\mathcal{G}M_{\rm halo}}{(a+R_{\rm halo})}
\Bigg[
1
-\frac{2\,r_{\rm ph0}}{3a}
\Bigg(
\left(2+\frac{a}{R_{\rm halo}}\right)  \\ & 
-2\left(1+\frac{R_{\rm halo}}{a}\right)\ln\!\Big(1+\frac{a}{R}\Big)
\Bigg)
\Bigg] \\&
+ \mathcal{O}\left(\frac{r_{\rm ph0}^{2}}{R_{\rm halo}^{2}},\frac{r_{\rm ph0}^{2}}{a^{2}}\right),\label{eq:deltaf_Her_halo}
\end{aligned}  
\end{equation}
%$\delta f (R_{\rm halo})\sim -\frac{2\mathcal{G} \delta M_{0}}{R}\left(1+ \frac{a}{R} \right)$.
and
\begin{equation}
\begin{aligned}
\delta g (R_{\rm halo})
\simeq &
\frac{2{\cal G}M_{\rm halo} R_{\rm halo}}{(a + R_{\rm halo})^{2}}
\left(1 + \frac{4\,r_{\rm ph0}}{3R_{\rm halo}}\right) \\ & +\mathcal{O}\left(\frac{r_{\rm ph0}^{2}}{R_{\rm halo}^{2}},\frac{r_{\rm ph0}^{2}}{a^{2}}\right).\label{eq:deltag_Her_halo}
\end{aligned}
\end{equation}
At the halo boundary, the perturbation $\delta f(R_{\rm halo})$ does not simply reduce
to $-2\mathcal{G}\,\delta m(R_{\rm halo})/R_{\rm halo}$, as the system is not yet in vacuum due to the \textit{ad hoc} definition of the halo radius. Instead, it matches the exterior gravitational potential generated by the full extended halo profile.
For the Hernquist distribution, this exterior solution corresponds to the
Newtonian potential
\begin{equation}
\Phi_{\rm H}(r)\simeq -\frac{2\mathcal{G}M_{\rm halo}}{r+a},
\end{equation}
up to corrections suppressed by $r_{\rm ph0}/a$.
This behavior is expected: while regularity requires the metric functions to be
continuous at $r=R_{\rm halo}$, the gravitational field outside the halo is
determined by the entire mass distribution rather than by the enclosed mass at
the matching radius alone. Consequently, at large radii, $r\gg R_{\rm halo}$, we recover the expected metric perturbations\footnote{In this regime, all contributions linear in $r_{\rm ph0}$ cancel. Consequently, we include the next-to-leading–order terms, unlike in Eqs.~(\ref{eq:deltaf_Her_halo}) and~(\ref{eq:deltag_Her_halo}), where the leading-order terms suffice.} 
\begin{equation}
\delta f(r)
\simeq
-\frac{2{\cal G}M_{\rm halo}}{r}
\left(1 - \frac{r_{\rm ph0}^2}{a^2}\right)+\mathcal{O}\!\left(\frac{r_{\rm ph0}^3}{a^3}\right),
\end{equation}
and
\begin{equation}
\delta g (r)
\simeq
\frac{2{\cal G}M_{\rm halo}}{r}
\left(1 - \frac{r_{\rm ph0}^2}{a^2}\right)+\mathcal{O}\!\left(\frac{r_{\rm ph0}^3}{a^3}\right),
\end{equation}
while in the asymptotic regime, $\delta f(r\to\infty)=\delta g(r\to\infty)=0$.  
The subleading term proportional to $r_{\rm ph0}^2/a^2$ encodes the finite-size
correction associated with the exclusion of the inner region
$r<r_{\rm ph0}$ and is parametrically suppressed for extended halos.

The perturbative consistency condition at radii $r\geq R_{\rm halo}$ is therefore set by
the ratio
\begin{equation}
r\geq R_{\rm halo}:\quad\frac{{\cal G}M_{\rm halo}}{r}\ll 1.
\end{equation}
For galaxy-scale dark matter halos, this condition is trivially satisfied.

%ensuring that the spacetime remains a small deformation of the Schwarzschild geometry even when the total halo mass is large.

\subsubsection{Near--photon--sphere limit}

While the perturbative deformation of the metric can be evaluated at arbitrary radii, the photon sphere constitutes the physically relevant scale for shadow observables. Focusing on this region, we first perturb Eq.~\eqref{eq:deltam_hernquist} around $r\simeq r_{\rm ph0}$, and then expand for $r_{\rm ph0}\ll a_s$, with $r_{s}=\frac{2}{3}r_{\rm ph0}$. The result gives
\begin{equation}
\begin{aligned}
\delta m(r)
\simeq &
\frac{M_{\rm halo}}{a^{2}}\Bigg[2\,r_{\rm ph0}\,(r-r_{\rm ph0}) 
+
(r-r_{\rm ph0})^{2} \left(1-\frac{6\,r_{\rm ph0}}{a} \right) \Bigg] \\&
+
\mathcal{O}\!\left(
(r-r_{\rm ph0})^{3},
\frac{r_{\rm ph0}^{2}}{a^{2}}
\right).
\label{eq:deltam_Her_nearph}
\end{aligned}
\end{equation}
Thus, the enclosed mass perturbation grows linearly with the radial displacement, indicating that the leading effect on the metric is governed by the local mass gradient of the halo rather than by its total mass. The enclosed mass vanishes effectively at the photon sphere.

Keeping next to leading contributions in the near photon sphere region
and discarding higher order terms in $r_{\rm ph0}/a$, the perturbation
$\delta g(r)$, given by Eq.~(\ref{eq:deltag_hernquist}), can be written in the compact form
\begin{equation}
\begin{aligned}
\delta g(r) \simeq &
\frac{36\,\mathcal{G}M_{\rm halo}}{a^{2}}
\Bigg[
\left(1-3\frac{r_{\rm ph0}}{a}\right)(r-r_{\rm ph0}) \\&
-\frac{3a}{2r_{\rm ph0}^2}
\left(1-\frac{8r_{\rm ph0}}{3a}\right)(r-r_{\rm ph0})^2
\Bigg]
\\[4pt]
&+\mathcal{O}\left(
(r-r_{\rm ph0})^{3},
\frac{r_{\rm ph0}^{2}}{a^{2}}
\right).
\end{aligned}
\label{eq:deltag-hernquist-ph}
\end{equation}
Similarly, the perturbation for $\delta f(r)$, Eq.~(\ref{eq:deltaf_hernquist}),
simplifies to
\begin{equation}
\begin{aligned}
\delta f(r)=&
-\frac{2\mathcal{G}M_{\rm halo}}{3a}
\Bigg[
\left(1-\frac{4r_{\rm ph0}}{a}\right) \left(1 + \frac{2}{r_{\rm ph0}} (r-r_{\rm ph0})\right)
\\[4pt]
&-\frac{2}{r_{\rm ph0}^2}
\left(1+\frac{r_{\rm ph0}}{2a}\right)(r-r_{\rm ph0})^2
\Bigg]
\\[4pt]
&+\mathcal{O}\left(
(r-r_{\rm ph0})^{3},
\frac{r_{\rm ph0}^{2}}{a^{2}}
\right).
\end{aligned}
\label{eq:deltaf-hernquist-ph}
\end{equation}
Thus, the leading term is controlled by the halo compactness
$\mathcal{C}={\cal G}M_{\rm halo}/a$, while the dominant relativistic correction
enters as a multiplicative factor $(1-4r_{\rm ph0}/a)$, encoding the effect of
the background curvature at the photon sphere. Higher order terms proportional to powers of $(r-r_{\rm ph0})$ do not contribute to shadow observables and are
therefore consistently neglected.

Evaluating the metric perturbation at the photon sphere, one
finds
\begin{equation}
r=r_{\rm ph0}:\;\; \delta f(r_{\rm ph0})\simeq-\frac{2\,\mathcal{G}M_{\rm halo}}{3a}
\left(1-\frac{4r_{\rm ph0}}{a}\right).\label{eq:deltaf-hernquist-ph-red}
\end{equation}

\subsubsection{Fractional deviation of the black hole shadow}

Using Eq.~(\ref{eqn:bcr_pert}) along with the expression for $\delta f(r)$ at the photon sphere, Eq.~(\ref{eq:deltaf-hernquist-ph-red}), the critical impact parameter is therefore
\begin{equation}
b_{\rm cr}
=
3\sqrt{3}\,{\cal G}M_0\left[1 +
\mathcal{C}
\left(
1-\frac{4 r_{\rm ph0}}{a}
\right)\right],
\label{eq:delta-q-hernquist-ph}
\end{equation}
which agrees at leading order with \cite{Cardoso:2021wlq}. The corresponding fractional deviation of the black hole shadow radius induced
by a Hernquist dark matter halo is therefore
\begin{equation}
\delta \simeq
\mathcal{C}
\left(
1-12\; \mathcal{C} \frac{M_{0}}{M_{\rm halo}}
\right).
%+\mathcal{O}\!\left(\mathcal{C}\frac{M_{0}}{M_{\rm halo}}\right)^2.
\end{equation}
Requiring consistency with current observational bounds on black hole shadow
sizes, $\delta\lesssim\mathcal{O}(0.1)$, leads to the conservative constraint
\begin{equation}
\mathcal{C} \lesssim 0.1,
\end{equation}
or equivalently,
\begin{equation}
a \gtrsim 10\,{\cal G} M_{\rm halo}.
\end{equation}
This result shows that black hole shadow observations primarily constrain the compactness of the Hernquist dark matter halo rather than its total mass
%implying that sufficiently extended halos can remain compatible with current measurements even when their total mass is large. 
A similar conclusion was reached in Ref.~\cite{Jha:2025xjf}.

\subsubsection{orbital-radius S2 star}

A natural question that arises in the vicinity of the black hole is whether the predicted enclosed dark matter mass remains consistent with current astronomical constraints from stellar dynamics.  Observations by the GRAVITY Collaboration probe the total enclosed mass
within the S2 orbital radius at the level \cite{GRAVITY:2020gka}
\begin{equation}
    \frac{\delta m (r_{\rm S2})}{M_{\rm BH,0}}\sim0.1\%.
\end{equation}
The enclosed mass in the inner region $r \ll a_s$, but still sufficiently far from the photon sphere
($r \gg r_{\rm ph0}$), follows the expected quadratic expansion
\begin{equation}
\delta m(r) \simeq M_{\rm halo} \frac{r^2}{a^{2}}.
\end{equation}
Evaluated at the S2 orbital radius (pericenter distance), this yields the simple estimate 
\begin{equation}
\frac{\delta m(r_{\rm S2})}{M_{\rm BH,0}}
\simeq
\frac{M_{\rm halo}}{M_{\rm BH,0}} \frac{r_{\rm S2}^2}{a^{2}}
\lesssim 10^{-3}.
\end{equation}
This inequality constrains the ratio
$M_{\rm halo}/a^{2}$. Taking parameters appropriate for Sgr~A$^{*}$, namely $M_{\rm BH,0}\simeq4.3\times10^{6}\,M_{\odot}$ and $r_{\rm S2}\simeq5.8\times10^{-4}\,\rm pc$, this constraint implies
\begin{equation}
\frac{M_{\rm halo}}{a^{2}}\;\lesssim\;1.3\times10^{10}\,M_{\odot}\,\mathrm{pc^{-2}}.
\end{equation}
For Milky Way–like dark matter haloes, analyses of rotation curves and dynamical models typically infer a cumulative mass of $M_{\rm vir}\sim10^{12}\,M_{\odot}$ with scale radii of order tens of kiloparsecs (e.g.\ $\sim10$–$20\,$kpc) \cite{2015ApJ...806...54E}, implying $M_{\rm halo}/a^{2}\sim 10^{4}\,M_{\odot}\,\mathrm{pc^{-2}}$, many orders of magnitude below the shadow bound. This demonstrates that, even for extended Hernquist haloes consistent with galaxy observations, the enclosed dark matter mass within the S2 orbit remains strongly subdominant. 
%supporting the validity of the perturbative scheme in the strong–field regime.

\subsection{Navarro--Frenk--White density profile}

We now consider a dark matter halo described by the Navarro--Frenk--White (NFW) density profile,
\begin{equation}
\rho(r)=\frac{\rho_s\,a_s}{r\left(1+r/a_s\right)^2},
\label{eq:NFW_rho}
\end{equation}
where $\rho_s$ is a characteristic density and $a_s$ denotes the scale radius. We deliberately use the notation $a_s$ to avoid confusion with the Schwarzschild radius $r_s$.

As before, the perturbative mass function $\delta m(r)$ is obtained by integrating the
density profile and imposing the inner boundary condition
$\delta m(r_{\rm ph0})=0$. From here, one finds the exact expression
\begin{equation}
\begin{aligned}
\delta m(r)
&=
4\pi a_s^3\rho_s
\Bigg[
\ln\!\left(\frac{a_s+r}{a_s+r_{\rm ph0}}\right)
\\
&\hspace{2.2em}
+
a_s\left(
\frac{1}{a_s+r}
-
\frac{1}{a_s+r_{\rm ph0}}
\right)
\Bigg].
\end{aligned}
\label{eq:deltam_nfw}
\end{equation}
This expression represents the general perturbative mass enclosed between the photon sphere and radius $r$, and fully accounts for the noncompact nature of the NFW halo.

The expression for $\delta f(r)$ can be obtained also in a closed form. This reads
\begin{equation}
\begin{aligned}
\delta f(r)
= &
-8\pi\,\mathcal{G}\,
a_s^3\rho_s
\Bigg[
-\frac{a_s}{a_s+r_{\rm ph0}}
-\ln(a_s+r_{\rm ph0})  \\ &
+\frac{a_s}{(a_s+r_s)} 
- \frac{a_s (r-r_s)}{(a_s+r_s)^{2}}
\ln\left(\frac{a_s+r}{r-r_s}\right) \\ &
+ \frac{(a_s+r)\ln(a_s+r)+(r_s-r)\ln(r-r_s)}{a_s+r_s}
\Bigg].
\end{aligned}
\label{eq:deltaf_nfw}
\end{equation}
Although some terms display apparent singular behavior near
$r=r_s$, the full expression for $\delta f(r)$ is regular for all $r>r_{\rm ph0}$, as required for the validity of the perturbative expansion. 
The expression for $\delta g(r)$ follows directly from its definition, Eq.~(\ref{eq:delta_g}), upon inserting the perturbative mass function $\delta m(r)$ given in Eq.~(\ref{eq:deltam_nfw}). One obtains 
\begin{equation}
\begin{aligned}
\delta g(r) &= \frac{8\pi\,\mathcal{G}\, a_s^3\rho_s}{r\,f_0(r)^2}
\Bigg[
\ln\!\left(\frac{a_s+r}{a_s+r_{\rm ph0}}\right)
\\
&\hspace{2.2em}
+
a_s\left(
\frac{1}{a_s+r}
-
\frac{1}{a_s+r_{\rm ph0}}
\right)
\Bigg]. 
\end{aligned}
\label{eq:deltag_nfw}
\end{equation}
which, together with $\delta f(r)$, fully characterizes the perturbative
deformation of the Schwarzschild geometry induced by an NFW dark matter halo. 

We now examine some physically relevant limiting regimes.

\subsubsection{Halo-edge limit}
At the outer edge of the halo, $r=R_{\rm halo}$, we assume  that $R_{\rm halo}\gg r_{\rm ph0}$, while keeping the ratio $R_{\rm halo}/a_s$
finite, as appropriate for realistic galactic halos. Expanding Eq.\
(\ref{eq:deltam_nfw}) in $r_{\rm ph0}/a_s\ll1$ up to second order, one finds 
\begin{equation}
\begin{aligned}
\delta M_0
\equiv
\delta m(R_{\rm halo})
&\simeq
4\pi a_s^3\rho_s
\Bigg[
\ln\!\left(1+\frac{R_{\rm halo}}{a_s}\right)
-
\frac{R_{\rm halo}}{a_s+R_{\rm halo}} \\
&\quad
- \frac{1}{2}\left(\frac{r_{\rm ph0}}{a_s}\right)^2
\Bigg] + \mathcal{O}\left(\frac{r_{\rm ph0}^3}{a_{s}^{3}} \right),
\end{aligned}
\label{eq:deltam_nfw_halo}
\end{equation}
where all linear terms in $r_{\rm ph0}$ cancel identically.
%as expected from the imposed boundary condition. Keeping terms up to , one finds
The last term inside the brackets accounts for the finite inner cutoff at the photon sphere and represents the leading correction due to the exclusion of the region $r<r_{\rm ph0}$. As in the Hernquist case, the inner cutoff correction also enters at second order in $r_{\rm ph0}/a_{s}$.

Expanding the perturbation $\delta g(r)$ from Eq.~(\ref{eq:deltag_nfw}), and making use of Eq.~(\ref{eq:deltam_nfw_halo}) together with the relation $r_s = 2r_{\rm ph0}/3$, the metric correction evaluated at the halo edge can be written in the compact form
\begin{equation}
\delta g(R)
=
\frac{2{\cal G}\,\delta M_0}{R_{\rm halo}}
\left(
1+\frac{4\,r_{\rm ph0}}{3R_{\rm halo}}
\right)
+
\mathcal{O}\!\left(
\frac{r_{\rm ph0}^2}{R_{\rm halo}^2},
\frac{r_{\rm ph0}^2}{a_s^2}
\right),
\label{eq:deltag_edge_expanded}
\end{equation}
where $\delta M_0$ is the enclosed dark matter mass within the halo radius $R_{\rm halo}$ (see Eq.~(\ref{eq:deltam_nfw_halo})).
Equation~\eqref{eq:deltag_edge_expanded} explicitly satisfies the defining relation given in Eq.~(\ref{eq:metric_edge}).

Expanding the perturbation $\delta f(r)$ from Eq.~(\ref{eq:deltaf_nfw}), one finds that all logarithmic dependence on the photon-sphere radius cancels. The resulting expression reads
\begin{equation}
\delta f(R)
=
-\,\frac{2\mathcal{G}\,\delta M_0}{R_{\rm halo}}
+
\mathcal{O}\!\left(
\frac{r_{\rm ph0}^2}{R_{\rm halo}^2},
\frac{r_{\rm ph0}^2}{a_s^2}
\right),
\label{eq:deltaf_edge_expanded}
\end{equation}
where $\delta M_0$ is given again by Eq.~(\ref{eq:deltam_nfw_halo}). 
Equation~\eqref{eq:deltaf_edge_expanded} confirms that, at the halo edge,
the metric function $f(r)$ depends only on the total enclosed mass, as
expected from the Schwarzschild form Eq.~(\ref{eq:metric_edge}).

It is worth noting that the hierarchy of correction orders is precisely what is expected in a weak-field expansion and ensures a consistent and well-controlled perturbative scheme. These expressions provide a transparent consistency check of the perturbative solution and of the matching between the interior black hole-halo geometry and the exterior vacuum Schwarzschild spacetime. 

%t is worth noting that the leading corrections to $\delta M_0$ and $\delta g(R)$ scale differently with the small parameter $r_{\rm ph0}/a_s$. While the enclosed mass perturbation behaves as $\delta M_0=\mathcal{O}\!\left((r_{\rm ph0}/a_s)^3\right)$, the correction to the metric function $g(r)$ at the halo edge is of order $\mathcal{O}\!\left((r_{\rm ph0}/a_s)^2\right)$. This apparent mismatch is not an inconsistency, but rather reflects the distinct geometric roles of the metric functions. The perturbation $\delta f(r)$ depends solely on the enclosed mass, whereas $\delta g(r)$ is sensitive to radial gradients of the mass function and includes the background factor $f_0^{-2}(r)$, whose expansion at $r=R$ introduces an additional power of $r_{\rm ph0}/R$. As a result, the hierarchy of correction orders is precisely what is expected in a weak-field expansion and ensures a consistent and well-controlled perturbative scheme.

\subsubsection{Near--photon--sphere limit}

We now consider the local behavior of the enclosed mass in the vicinity of the photon sphere, $r\simeq r_{\rm ph0}$. Expanding
Eq.~\eqref{eq:deltam_nfw} for the hierarchy assumption $r_{\rm ph0}\ll a_s$, and considering $r_{s}=\frac{2}{3}r_{\rm ph0}$, one finds to leading order
\begin{equation}
\begin{aligned}
\delta m(r)
 \simeq  &
2\pi a_s\rho_s \Bigg[2r_{\rm rph0} (r-r_{\rm rph0})+
\left(1 - \frac{4r_{\rm ph0}}{a_{s}} \right) \bigl(r-r_{\rm ph0}\bigr)^2\Bigg]  \\
& +  \mathcal{O}\!\left(
(r-r_{\rm ph0})^3,
\frac{r_{\rm ph0}^{2}}{a_s^{2}}
\right).
\label{eq:deltam_nfw_nearph}
\end{aligned}
\end{equation}
This expression satisfies the imposed boundary condition
$\delta m(r_{\rm ph0})=0$ and shows that, in the vicinity of the photon sphere, the enclosed dark matter mass increases linearly with the radial displacement. This behavior is the local manifestation of the underlying $r^{2}$ scaling implied by the inner cusp $\rho\propto r^{-1}$.
%\footnote{ This follows directly from the expansion $r^{2}-r_{\rm ph0}^{2} \simeq2r_{\rm ph0}(r-r_{\rm ph0})+(r-r_{\rm ph0})^{2}$.}. 
Despite the formally divergent total mass of the NFW profile at large radii, the local enclosed mass in the strong--field region remains finite and parametrically small for $r\ll a_s$. 

Expanding the perturbed metric function in Eq.~(\ref{eq:deltag_nfw}) around the photon sphere, and adopting the same hierarchy assumptions as before, one obtains
\begin{equation}
\begin{aligned}
\delta g(r)
\simeq{}&
72\pi\,\mathcal{G}\,a_s\rho_s
\left(1 - 2\frac{r_{\rm ph0}}{a_s}\right)
\,(r-r_{\rm ph0})
\\[0.4em]
&\quad
-36\pi\,\mathcal{G}\,\frac{a_s\rho_s}{r_{\rm ph0}}
\left(9 - 16 \frac{r_{\rm ph0}}{a_s}\right)
\,(r-r_{\rm ph0})^2
\\[0.4em]
&\quad
+ \mathcal{O}\!\left(
(r-r_{\rm ph0})^3,
\frac{r_{\rm ph0}^{2}}{a_s^{2}}
\right).
\end{aligned}
\label{eq:deltag_nfw_photon_expanded}
\end{equation}
In deriving this expression, we have consistently neglected subleading
contributions suppressed by additional powers of $r_{\rm ph0}/a_s$. The above expression explicitly satisfies the imposed
boundary condition $\delta g(r_{\rm ph0})=0$, ensuring regularity of the
perturbed geometry at the photon sphere. The leading correction grows
linearly with the radial displacement $(r-r_{\rm ph0})$, while the ratios $r_{\rm ph0}/a_{s}$ control the hierarchy scale at first order.

As a nontrivial consistency check, one may verify that the expansions
in Eqs.~(\ref{eq:deltam_nfw_nearph}) and~(\ref{eq:deltag_nfw_photon_expanded})
satisfy the defining relation
$\delta g(r)=\frac{2\mathcal{G}\,\delta m(r)}{r\,f_0(r)^2}$
in the vicinity of the photon sphere\footnote{
This relation is an exact identity at the level of the first order expansion. However, once $\delta m(r)$ and $f_0(r)^{-2}$ are independently expanded and truncated at some order, the identity need not be satisfied order by order. Subleading terms discarded in $\delta m$ or in $f_0^{-2}$ can combine to contribute at the same perturbative order in $\delta g$, leading to different quadratic coefficients. The leading (linear) behavior is scheme-independent, while higher-order terms are intrinsically sensitive to the chosen truncation.}.
%In performing this check, it is essential to retain a consistent order in both $\delta m(r)$ and the expansion of $f_0(r)^2$. Including or neglecting  ubleading terms inconsistently may introduce spurious second--order contributions, leading to an apparent mismatch in the quadratic coefficient of $\delta g(r)$. Hence, the precise coefficient of the quadratic correction depends on how subleading terms are consistently truncated.

Expanding the metric perturbation $\delta f(r)$, Eq.~(\ref{eq:deltaf_nfw}), near the photon sphere and retaining the leading powers of the scale radius $a_s$, while neglecting subleading terms suppressed by ratios $r_{\rm ph0}/a_s$, one finds that the dominant contributions take the compact form
\begin{equation}
\begin{aligned}
\delta f(r)
\simeq & - \frac{8\pi\,\mathcal{G}\,\rho_s\,a_s^2}{3}
\Bigg[
1 +\frac{2(r-r_{\rm ph0})}{3r_{\rm ph0}}\left(3-\frac{5 r_{\rm ph0}}{a_{s}} \right)
\Bigg] \\&
+
 \mathcal{O}\!\left(
(r-r_{\rm ph0})^2,
\frac{r_{\rm ph0}^{2}}{a_s^{2}}
\right).
\label{eq:deltaf_photon_sphere}
\end{aligned}
\end{equation}
%The ratio $r/r_{\rm rph0}$ further ensures that the metric deformation remains perturbatively controlled in the strong--field region, consistently justifying the use of unmodified NFW profiles in the analysis of shadow observables.
Unlike the perturbation $\delta g(r)$, $\delta f(r)$ does not
vanish at the photon sphere radius,
yielding
\begin{equation}
r=r_{\rm ph0}:\;\; \delta f(r_{\rm ph0})\simeq-8\pi\mathcal{G}\rho_s a_s^2/3.
\end{equation}
This finite offset directly contributes to the black hole shadow size and, unlike the Hernquist contribution (see Eq.~(\ref{eq:deltaf-hernquist-ph-red})), does not involve factors of the form $\frac{r_{\rm ph0}}{a}$.

\subsubsection{Fractional deviation of the black hole shadow}

The critical impact parameter is therefore
\begin{equation}
b_{\rm cr}
=
3\sqrt{3}\,{\cal G}M_0\left[1 +
4\pi\mathcal{G}\rho_s a_s^2\right].
\label{eq:delta-b-NFW-ph}
\end{equation}
From here, we can identify the fractional deviation of the black hole shadow
\begin{equation}
    \delta \simeq 4\pi\mathcal{G}\rho_s a_s^2.
    %+ \mathcal{O}\!\left(\frac{r_{\rm ph0}}{a_s},\frac{r}{a_s}\right).
\end{equation}
Importantly, this result depends only on the local combination
$\rho_s a_s^2$, which control the inner structure of the density profile, and it is insensitive to the formally divergent total mass of the
NFW halo. 

Requiring consistency with current observational bounds on black hole shadow sizes,
$\delta \lesssim \mathcal{O}(0.1)$, leads to the constraint
\begin{equation}
\rho_s\,a_s^2
\;\lesssim\;
\frac{0.1\,c^2}{4\pi\,\mathcal{G}}
\;\sim\;
10^{11}\,M_\odot\,\mathrm{pc^{-1}} .
\end{equation}
This bound constrains the inner normalization of the NFW profile, which controls the compactness of the dark matter
distribution in the strong--field region. 

For Milky Way–like halos, fits to rotation curves and dynamical models using NFW
profiles typically yield characteristic parameters such as a scale radius
$a_s\sim10\!-\!20\ {\rm kpc}$ with a characteristic density
$\rho_s\sim(0.006\!-\!0.01)\,M_\odot\,{\rm pc^{-3}}$ (e.g.,
\cite{Sofue:2013kja}). From these values, one estimates
\begin{equation}
\rho_s a_s^2
\;\sim\; 10^{6}\,
M_\odot\,\mathrm{pc^{-1}} .
\end{equation}
This is roughly five orders of magnitude below the conservative bound inferred from black hole shadow observations. 
%Therefore, standard Milky Way--like NFW halos lie deeply within the perturbative regime required for a controlled deformation of the Schwarzschild geometry. 
This comparison highlights that black hole shadow measurements are sensitive only to extremely compact dark matter configurations, while
realistic galactic halos produce very small strong--field effects.

\subsubsection{orbital-radius S2 star}

As before, we evaluate whether the predicted enclosed dark matter mass remains consistent with the current astronomical constraint from stellar dynamics.  In the inner region $r \ll a_s$, but still sufficiently far from the photon sphere
($r \gg r_{\rm ph0}$), the enclosed mass in Eq.~(\ref{eq:deltam_nfw}) admits the expansion
\begin{equation}
\delta m(r) \simeq 2\pi\,a_s\,\rho_s\, r^2.
\end{equation}
Evaluated at the S2 orbital radius gives the estimate 
\begin{equation}
\frac{\delta m(r_{\rm S2})}{M_{\rm BH,0}}
\simeq
2\pi\,\frac{a_s\,\rho_s}{M_{\rm BH,0}}\, r_{\rm S2}^2
\lesssim 10^{-3}.
\end{equation}
This inequality constrains the product $a_s\rho_s$, which can be interpreted as
an effective surface density governing the inner normalization of the NFW
profile. Interestingly, this bound is local and independent of the total halo mass,
which may be arbitrarily large.

Adopting parameters appropriate for Sgr~A$^{*}$,
$M_{\rm BH,0}\simeq4.3\times10^{6}M_\odot$ and
$r_{\rm S2}\simeq5.8\times10^{-4},{\rm pc}$ \cite{GRAVITY:2020gka},
the observational constraint translates into the numerical bound
\begin{equation}
a_s\rho_s \lesssim 2\times10^{9}
M_\odot\,{\rm pc^{-2}} .
\end{equation}

For Milky Way–like halos, typical parameters are 
$a_s\sim10\!-\!20\ {\rm kpc}$ and 
$\rho_s\sim(0.006\!-\!0.01)\,M_\odot\,{\rm pc^{-3}}$ (e.g.,
\cite{Sofue:2013kja}), yielding 
\begin{equation}
    a_s\rho_s\sim10-100\,M_{\odot}\;{\rm pc^{-2}},
\end{equation}
many orders of magnitude below this limit. Hence, 
despite the formally divergent mass of the NFW profile at large radii, the enclosed dark matter mass within stellar orbits remains finite and parametrically small. This further confirms that stellar observables probe only the inner structure of the dark matter halo.

%We therefore conclude that NFW--type dark matter halos naturally satisfy the conditions required for a controlled perturbative deformation of the Schwarzschild geometry in the strong--field regime, making them particularly suitable for assessing dark matter imprints on black hole shadow observables and stellar dynamics.

%%%%%%%%%%%%%%%%%%%%%%%%%%%%%%%%%%%%%%%%%%%%%%%%%%%%%%%%%%%%%%%%
\section{Conclusions}\label{sec:Conclusions}
%%%%%%%%%%%%%%%%%%%%%%%%%%%%%%%%%%%%%%%%%%%%%%%%%%%%%%%%%%%%%%%%

%This work builds upon previous studies of metric deformations induced by dark matter distributions around black holes, contributing to the broader effort to constrain dark matter properties in the strong-field regime through shadow observables. Our starting point is the perturbative framework introduced in Ref.~\cite{Gomez:2024ack}. Here, we revisit and clarify key conceptual aspects of this approach and extend its application to physically motivated galactic halo profiles.

Using the perturbative framework introduced in Ref.~\cite{Gomez:2024ack}, we have derived the perturbed metric functions induced by surrounding dark matter distributions at all relevant radii. This construction consistently captures both the strong–field region near the black hole and the weak–field behavior at large distances, while preserving asymptotic flatness. Within this setup, we have also obtained general, model–independent expressions for the photon sphere radius and for the black hole shadow in any static, spherically symmetric geometry, thereby providing a systematic way to relate dark matter–induced metric deformations to observable quantities. In particular, we have examined the validity of the perturbative expansion at the photon–sphere scale and near the halo boundary, assuming a natural hierarchy of scales determined by the black hole and halo parameters.

%Although the total mass of a dark matter halo typically exceeds the black hole mass by many orders of magnitude, the perturbative expansion remains well controlled because only an extremely small fraction of the halo mass is enclosed within the photon sphere. For extended density profiles such as the Hernquist and NFW halos, the ratio $\delta m(r_{\rm ph0})/M_0 \ll 1$ is exceedingly small, ensuring the validity of the perturbative treatment in the region probed by shadow observables.

We found that deformations of the shadow radius are determined by local properties of the halo. For the Hernquist profile, the corrections are governed by the effective compactness of the dark matter distribution, while for the NFW profile they are controlled by the characteristic combination $\rho_s a_s^2$. These results show that extended dark matter halos induce finite and well-behaved modifications of the spacetime geometry in the strong–field region. In agreement with previous studies and current expectations, the corresponding corrections to the shadow radius remain well within existing observational bounds. As an additional consistency check, we have estimated the total dark matter mass enclosed within the orbital radius of the S2 star, finding values safely below the $0.1\%$ upper limit reported by the GRAVITY collaboration.

While the present analysis can be readily extended to any static, spherically symmetric black hole and to a broad class of dark matter density profiles, several important directions remain to be explored. A natural next step is the investigation of more compact and self–gravitating dark matter configurations surrounding black holes,
such as dense spikes or ultracompact halos, whose gravitational backreaction may produce stronger and potentially observable deviations from vacuum geometries \cite{Sadeghian:2013laa,Speeney:2022ryg,Bertone:2024wbn}. 
%In such scenarios, the perturbative framework developed here can serve as a first step toward identifying the parameter space where nonlinear effects become relevant and fully relativistic treatments are required.

Another key extension concerns rotating black hole spacetimes. Generalizing the present scheme to axisymmetric backgrounds embedded in dark matter environments would provide a more realistic description of these systems (see e.g. \cite{Fernandes:2025osu,Datta:2026krm}).
%where spin plays a central role in determining shadow morphology, accretion dynamics, and gravitational wave emission. 
In this context, the perturbative approach could offer valuable analytical insight into how dark matter environments modify geodesic structure, photon rings, and quasinormal mode spectra in Kerr–like geometries.

More broadly, this framework is particularly well suited for studying dark matter effects in extreme mass–ratio inspiral (EMRI) systems, which are among the primary targets of the future space–based LISA interferometer. In such environments, dark matter distributions can induce small but cumulative corrections to orbital evolution leading to gravitational wave phase shifts (see e.g \cite{Kavanagh:2020cfn,Speeney:2022ryg} and references therein). The
perturbative scheme developed here provides a systematic and computationally efficient tool to quantify these effects, thereby contributing to the broader effort of probing dark matter properties through multi–messenger observations of black holes.

%\section*{Acknowledgments}

%\bibliographystyle{alpha}
\bibliography{biblio}

@article{Konoplya:2019sns,
    author = "Konoplya, R. A.",
    title = "{Shadow of a black hole surrounded by dark matter}",
    eprint = "1905.00064",
    archivePrefix = "arXiv",
    primaryClass = "gr-qc",
    doi = "10.1016/j.physletb.2019.05.043",
    journal = "Phys. Lett. B",
    volume = "795",
    pages = "1--6",
    year = "2019"
}

@article{Konoplya:2022hbl,
    author = "Konoplya, R. A. and Zhidenko, A.",
    title = "{Solutions of the Einstein Equations for a Black Hole Surrounded by a Galactic Halo}",
    eprint = "2202.02205",
    archivePrefix = "arXiv",
    primaryClass = "gr-qc",
    doi = "10.3847/1538-4357/ac76bc",
    journal = "Astrophys. J.",
    volume = "933",
    number = "2",
    pages = "166",
    year = "2022"
}

@article{Macedo:2024qky,
    author = "Macedo, Caio F. B. and Rosa, Jo\~ao Lu\'\i{}s and Rubiera-Garcia, Diego",
    title = "{Optical appearance of black holes surrounded by a dark matter halo}",
    eprint = "2402.13047",
    archivePrefix = "arXiv",
    primaryClass = "gr-qc",
    month = "2",
    year = "2024"
}

@article{Cardoso:2021wlq,
    author = "Cardoso, Vitor and Destounis, Kyriakos and Duque, Francisco and Macedo, Rodrigo Panosso and Maselli, Andrea",
    title = "{Black holes in galaxies: Environmental impact on gravitational-wave generation and propagation}",
    eprint = "2109.00005",
    archivePrefix = "arXiv",
    primaryClass = "gr-qc",
    doi = "10.1103/PhysRevD.105.L061501",
    journal = "Phys. Rev. D",
    volume = "105",
    number = "6",
    pages = "L061501",
    year = "2022"
}

@article{EventHorizonTelescope:2022xqj,
    author = "Akiyama, Kazunori and others",
    collaboration = "Event Horizon Telescope",
    title = "{First Sagittarius A* Event Horizon Telescope Results. VI. Testing the Black Hole Metric}",
    eprint = "2311.09484",
    archivePrefix = "arXiv",
    primaryClass = "astro-ph.HE",
    reportNumber = "FERMILAB-PUB-22-422-PPD",
    doi = "10.3847/2041-8213/ac6756",
    journal = "Astrophys. J. Lett.",
    volume = "930",
    number = "2",
    pages = "L17",
    year = "2022"
}

@article{Gondolo:1999ef,
    author = "Gondolo, Paolo and Silk, Joseph",
    title = "{Dark matter annihilation at the galactic center}",
    eprint = "astro-ph/9906391",
    archivePrefix = "arXiv",
    reportNumber = "MPI-PHT-99-10, OUAST-99-9",
    doi = "10.1103/PhysRevLett.83.1719",
    journal = "Phys. Rev. Lett.",
    volume = "83",
    pages = "1719--1722",
    year = "1999"
}

@article{Sadeghian:2013laa,
    author = "Sadeghian, Laleh and Ferrer, Francesc and Will, Clifford M.",
    title = "{Dark matter distributions around massive black holes: A general relativistic analysis}",
    eprint = "1305.2619",
    archivePrefix = "arXiv",
    primaryClass = "astro-ph.GA",
    doi = "10.1103/PhysRevD.88.063522",
    journal = "Phys. Rev. D",
    volume = "88",
    number = "6",
    pages = "063522",
    year = "2013"
}

@article{Hernquist:1990be,
    author = "Hernquist, Lars",
    title = "{An Analytical Model for Spherical Galaxies and Bulges}",
    reportNumber = "IASSNS-AST-89-63",
    doi = "10.1086/168845",
    journal = "Astrophys. J.",
    volume = "356",
    pages = "359",
    year = "1990"
}

@article{Abbott:2016blz,
    author = "Abbott, B.P. and others",
    collaboration = "LIGO Scientific, Virgo",
    title = "{Observation of Gravitational Waves from a Binary Black Hole Merger}",
    eprint = "1602.03837",
    archivePrefix = "arXiv",
    primaryClass = "gr-qc",
    reportNumber = "LIGO-P150914",
    doi = "10.1103/PhysRevLett.116.061102",
    journal = "Phys. Rev. Lett.",
    volume = "116",
    number = "6",
    pages = "061102",
    year = "2016"
}

@article{Kavanagh:2020cfn,
    author = "Kavanagh, Bradley J. and Nichols, David A. and Bertone, Gianfranco and Gaggero, Daniele",
    title = "{Detecting dark matter around black holes with gravitational waves: Effects of dark-matter dynamics on the gravitational waveform}",
    eprint = "2002.12811",
    archivePrefix = "arXiv",
    primaryClass = "gr-qc",
    month = "2",
    year = "2020"
}

@ARTICLE{2022ApJ...930L..17E,
       author = {{Event Horizon Telescope Collaboration} and {Akiyama}, Kazunori and {Alberdi}, Antxon and {Alef}, Walter and {Algaba}, Juan Carlos and {Anantua}, Richard and {Asada}, Keiichi and {Azulay}, Rebecca and {Bach}, Uwe and {Baczko}, Anne-Kathrin and {Ball}, David and {Balokovi{\'c}}, Mislav and {Barrett}},
        title = "{First Sagittarius A* Event Horizon Telescope Results. VI. Testing the Black Hole Metric}",
      journal = {\apjl},
         year = 2022,
        month = may,
       volume = {930},
       number = {2},
          eid = {L17},
        pages = {L17},
          doi = {10.3847/2041-8213/ac6756},
       adsurl = {https://ui.adsabs.harvard.edu/abs/2022ApJ...930L..17E}
}

@article{EventHorizonTelescope:2019dse,
    author = "Akiyama, Kazunori and others",
    collaboration = "Event Horizon Telescope",
    title = "{First M87 Event Horizon Telescope Results. I. The Shadow of the Supermassive Black Hole}",
    eprint = "1906.11238",
    archivePrefix = "arXiv",
    primaryClass = "astro-ph.GA",
    doi = "10.3847/2041-8213/ab0ec7",
    journal = "Astrophys. J. Lett.",
    volume = "875",
    pages = "L1",
    year = "2019"
}

@article{EventHorizonTelescope:2022wkp,
    author = "Akiyama, Kazunori and others",
    collaboration = "Event Horizon Telescope",
    title = "{First Sagittarius A* Event Horizon Telescope Results. I. The Shadow of the Supermassive Black Hole in the Center of the Milky Way}",
    doi = "10.3847/2041-8213/ac6674",
    journal = "Astrophys. J. Lett.",
    volume = "930",
    number = "2",
    pages = "L12",
    year = "2022"
}

@article{GRAVITY:2020gka,
    author = "Abuter, R. and others",
    collaboration = "GRAVITY",
    title = "{Detection of the Schwarzschild precession in the orbit of the star S2 near the Galactic centre massive black hole}",
    eprint = "2004.07187",
    archivePrefix = "arXiv",
    primaryClass = "astro-ph.GA",
    doi = "10.1051/0004-6361/202037813",
    journal = "Astron. Astrophys.",
    volume = "636",
    pages = "L5",
    year = "2020"
}

@article{Babichev:2012sg,
    author = "Babichev, E. and Dokuchaev, V. and Eroshenko, Yu.",
    title = "{Backreaction of accreting matter onto a black hole in the Eddington-Finkelstein coordinates}",
    eprint = "1202.2836",
    archivePrefix = "arXiv",
    primaryClass = "gr-qc",
    doi = "10.1088/0264-9381/29/11/115002",
    journal = "Class. Quant. Grav.",
    volume = "29",
    pages = "115002",
    year = "2012"
}

@article{Xu:2018wow,
    author = "Xu, Zhaoyi and Hou, Xian and Gong, Xiaobo and Wang, Jiancheng",
    title = "{Black Hole Space-time In Dark Matter Halo}",
    eprint = "1803.00767",
    archivePrefix = "arXiv",
    primaryClass = "gr-qc",
    doi = "10.1088/1475-7516/2018/09/038",
    journal = "JCAP",
    volume = "09",
    pages = "038",
    year = "2018"
}

@article{Saurabh:2020zqg,
    author = "Saurabh, K. and Jusufi, Kimet",
    title = "{Imprints of dark matter on black hole shadows using spherical accretions}",
    eprint = "2009.10599",
    archivePrefix = "arXiv",
    primaryClass = "gr-qc",
    doi = "10.1140/epjc/s10052-021-09280-9",
    journal = "Eur. Phys. J. C",
    volume = "81",
    number = "6",
    pages = "490",
    year = "2021"
}

@article{Anjum:2023axh,
    author = "Anjum, Arshia and Afrin, Misba and Ghosh, Sushant G.",
    title = "{Investigating effects of dark matter on photon orbits and black hole shadows}",
    eprint = "2301.06373",
    archivePrefix = "arXiv",
    primaryClass = "gr-qc",
    doi = "10.1016/j.dark.2023.101195",
    journal = "Phys. Dark Univ.",
    volume = "40",
    pages = "101195",
    year = "2023"
}

@article{Psaltis:2007rv,
    author = "Psaltis, Dimitrios",
    title = "{Testing General Metric Theories of Gravity with Bursting Neutron Stars}",
    eprint = "0704.2426",
    archivePrefix = "arXiv",
    primaryClass = "astro-ph",
    doi = "10.1103/PhysRevD.77.064006",
    journal = "Phys. Rev. D",
    volume = "77",
    pages = "064006",
    year = "2008"
}

@article{EventHorizonTelescope:2020qrl,
    author = "Psaltis, Dimitrios and others",
    collaboration = "Event Horizon Telescope",
    title = "{Gravitational Test Beyond the First Post-Newtonian Order with the Shadow of the M87 Black Hole}",
    eprint = "2010.01055",
    archivePrefix = "arXiv",
    primaryClass = "gr-qc",
    doi = "10.1103/PhysRevLett.125.141104",
    journal = "Phys. Rev. Lett.",
    volume = "125",
    number = "14",
    pages = "141104",
    year = "2020"
}

@article{Volkel:2020xlc,
    author = {V{\"o}lkel, Sebastian H. and Barausse, Enrico and Franchini, Nicola and Broderick, Avery E.},
    title = "{EHT tests of the strong-field regime of general relativity}",
    eprint = "2011.06812",
    archivePrefix = "arXiv",
    primaryClass = "gr-qc",
    doi = "10.1088/1361-6382/ac27ed",
    journal = "Class. Quant. Grav.",
    volume = "38",
    number = "21",
    pages = "21LT01",
    year = "2021"
}

@article{Bardeen:1973tla,
    author = "Bardeen, J. M.",
    editor = "DeWitt, C{\'e}cile and DeWitt, Bryce Seligman",
    title = "{Timelike and null geodesics in the Kerr metric}",
    journal = "Proceedings, Ecole d'Et{\'e} de Physique Th{\'e}orique: Les Astres Occlus : Les Houches, France, August, 1972, 215-240",
    pages = "215--240",
    year = "1973"
}

@article{Perlick:2021aok,
    author = "Perlick, Volker and Tsupko, Oleg Yu.",
    title = "{Calculating black hole shadows: Review of analytical studies}",
    eprint = "2105.07101",
    archivePrefix = "arXiv",
    primaryClass = "gr-qc",
    doi = "10.1016/j.physrep.2021.10.004",
    journal = "Phys. Rept.",
    volume = "947",
    pages = "1--39",
    year = "2022"
}

@article{LIGOScientific:2017ync,
    author = "Abbott, B. P. and others",
    collaboration = "LIGO Scientific, Virgo, Fermi GBM, INTEGRAL, IceCube, AstroSat Cadmium Zinc Telluride Imager Team, IPN, Insight-Hxmt, ANTARES, Swift, AGILE Team, 1M2H Team, Dark Energy Camera GW-EM, DES, DLT40, GRAWITA, Fermi-LAT, ATCA, ASKAP, Las Cumbres Observatory Group, OzGrav, DWF (Deeper Wider Faster Program), AST3, CAASTRO, VINROUGE, MASTER, J-GEM, GROWTH, JAGWAR, CaltechNRAO, TTU-NRAO, NuSTAR, Pan-STARRS, MAXI Team, TZAC Consortium, KU, Nordic Optical Telescope, ePESSTO, GROND, Texas Tech University, SALT Group, TOROS, BOOTES, MWA, CALET, IKI-GW Follow-up, H.E.S.S., LOFAR, LWA, HAWC, Pierre Auger, ALMA, Euro VLBI Team, Pi of Sky, Chandra Team at McGill University, DFN, ATLAS Telescopes, High Time Resolution Universe Survey, RIMAS, RATIR, SKA South Africa/MeerKAT",
    title = "{Multi-messenger Observations of a Binary Neutron Star Merger}",
    eprint = "1710.05833",
    archivePrefix = "arXiv",
    primaryClass = "astro-ph.HE",
    reportNumber = "LIGO-P1700294, VIR-0802A-17, FERMILAB-PUB-17-478-A-AE-CD",
    doi = "10.3847/2041-8213/aa91c9",
    journal = "Astrophys. J. Lett.",
    volume = "848",
    number = "2",
    pages = "L12",
    year = "2017"
}

@article{Sofue:2013kja,
    author = "Sofue, Yoshiaki",
    title = "{Rotation Curve and Mass Distribution in the Galactic Center --- From Black Hole to Entire Galaxy ---}",
    eprint = "1307.8241",
    archivePrefix = "arXiv",
    primaryClass = "astro-ph.GA",
    doi = "10.1093/pasj/65.6.118",
    journal = "Publ. Astron. Soc. Jap.",
    volume = "65",
    pages = "118",
    year = "2013"
}

@article{Springel:2004kf,
    author = "Springel, Volker and Di Matteo, Tiziana and Hernquist, Lars",
    title = "{Modeling feedback from stars and black holes in galaxy mergers}",
    eprint = "astro-ph/0411108",
    archivePrefix = "arXiv",
    doi = "10.1111/j.1365-2966.2005.09238.x",
    journal = "Mon. Not. Roy. Astron. Soc.",
    volume = "361",
    pages = "776--794",
    year = "2005"
}

@article{Kamermans:2024ieb,
    author = "Kamermans, Jasper Leonora P. D. and Wierda, A. Renske A. C.",
    title = "{Darkness visible: N-body simulations of dark matter spikes in Hernquist haloes}",
    eprint = "2411.12007",
    archivePrefix = "arXiv",
    primaryClass = "astro-ph.CO",
    doi = "10.1093/mnras/staf517",
    journal = "Mon. Not. Roy. Astron. Soc.",
    volume = "539",
    number = "1",
    pages = "135--144",
    year = "2025"
}

@article{Chakraborty:2024gcr,
    author = "Chakraborty, Sumanta and Comp{\`e}re, Geoffrey and Machet, Ludovico",
    title = "{Tidal Love numbers and quasinormal modes of the Schwarzschild-Hernquist black hole}",
    eprint = "2412.14831",
    archivePrefix = "arXiv",
    primaryClass = "gr-qc",
    doi = "10.1103/4p2c-rwdh",
    journal = "Phys. Rev. D",
    volume = "112",
    number = "2",
    pages = "024015",
    year = "2025"
}

@ARTICLE{2015ApJ...806...54E,
       author = {{Eadie}, Gwendolyn M. and {Harris}, William E. and {Widrow}, Lawrence M.},
        title = "{Estimating the Galactic Mass Profile in the Presence of Incomplete Data}",
      journal = {\apj},
         year = 2015,
        month = jun,
       volume = {806},
       number = {1},
          eid = {54},
        pages = {54},
          doi = {10.1088/0004-637X/806/1/54},
archivePrefix = {arXiv},
       eprint = {1503.07176},
 primaryClass = {astro-ph.GA},
       adsurl = {https://ui.adsabs.harvard.edu/abs/2015ApJ...806...54E}
}

@article{Jha:2025xjf,
    author = "Jha, Sohan Kumar",
    title = "{Thermodynamics, weak gravitational lensing, and parameter estimation of a Schwarzschild black hole immersed in Hernquist dark matter halo}",
    eprint = "2503.19938",
    archivePrefix = "arXiv",
    primaryClass = "gr-qc",
    doi = "10.1088/1475-7516/2025/06/033",
    journal = "JCAP",
    volume = "06",
    pages = "033",
    year = "2025"
}

@article{Gomez:2024ack,
    author = "G{\'o}mez, Gabriel and Valageas, Patrick",
    title = "{Constraining self-interacting scalar field dark matter from the black hole shadow of the Event Horizon Telescope}",
    eprint = "2403.08988",
    archivePrefix = "arXiv",
    primaryClass = "astro-ph.CO",
    doi = "10.1103/PhysRevD.109.103038",
    journal = "Phys. Rev. D",
    volume = "109",
    number = "10",
    pages = "103038",
    year = "2024"
}

@article{Zhao:2026yis,
    author = "Zhao, Yang and Gong, Yungui",
    title = "{Dark matter distributions around extreme mass ratio inspirals: effects of radial pressure and relativistic treatment}",
    eprint = "2602.12022",
    archivePrefix = "arXiv",
    primaryClass = "gr-qc",
    month = "2",
    year = "2026"
}

@article{Datta:2023zmd,
    author = "Datta, Sayak",
    title = "{Black holes immersed in dark matter: Energy condition and sound speed}",
    eprint = "2312.01277",
    archivePrefix = "arXiv",
    primaryClass = "gr-qc",
    doi = "10.1103/PhysRevD.109.104042",
    journal = "Phys. Rev. D",
    volume = "109",
    number = "10",
    pages = "104042",
    year = "2024"
}

@article{Fernandes:2025osu,
    author = "Fernandes, Pedro G. S. and Cardoso, Vitor",
    title = "{Spinning Black Holes in Astrophysical Environments}",
    eprint = "2507.04389",
    archivePrefix = "arXiv",
    primaryClass = "gr-qc",
    doi = "10.1103/9shv-5d21",
    journal = "Phys. Rev. Lett.",
    volume = "135",
    number = "21",
    pages = "211403",
    year = "2025"
}

@article{Datta:2026krm,
    author = "Datta, Sayak and Singha, Chiranjeeb",
    title = "{Geometric properties of slowly rotating black holes embedded in matter environments}",
    eprint = "2602.10579",
    archivePrefix = "arXiv",
    primaryClass = "gr-qc",
    month = "2",
    year = "2026"
}

@article{Fauzi:2025yse,
    author = "Fauzi, M. F. and Ramadhan, H. S. and Sulaksono, A.",
    title = "{Two descriptions of dark matter around a black hole: photon sphere, shadow, and lensing}",
    eprint = "2512.17304",
    archivePrefix = "arXiv",
    primaryClass = "gr-qc",
    month = "12",
    year = "2025"
}

@article{Speeney:2022ryg,
    author = "Speeney, Nicholas and Antonelli, Andrea and Baibhav, Vishal and Berti, Emanuele",
    title = "{Impact of relativistic corrections on the detectability of dark-matter spikes with gravitational waves}",
    eprint = "2204.12508",
    archivePrefix = "arXiv",
    primaryClass = "gr-qc",
    doi = "10.1103/PhysRevD.106.044027",
    journal = "Phys. Rev. D",
    volume = "106",
    number = "4",
    pages = "044027",
    year = "2022"
}

@article{Bolokhov:2025zva,
    author = "Bolokhov, S. V.",
    title = "{Revisiting black holes in dark-matter halos: on consistent solutions to the Einstein equations}",
    eprint = "2512.06930",
    archivePrefix = "arXiv",
    primaryClass = "gr-qc",
    month = "12",
    year = "2025"
}

@article{Bertone:2024wbn,
    author = "Bertone, Gianfranco and Wierda, A. Renske A. C. and Gaggero, Daniele and Kavanagh, Bradley J. and Volonteri, Marta and Yoshida, Naoki",
    title = "{Toward a realistic description of dark matter overdensities around black holes}",
    eprint = "2404.08731",
    archivePrefix = "arXiv",
    primaryClass = "astro-ph.CO",
    doi = "10.1103/5nnf-8fz9",
    journal = "Phys. Rev. D",
    volume = "112",
    number = "4",
    pages = "043537",
    year = "2025"
}

@article{Matos:2003nb,
    author = "Matos, Tonatiuh and Nunez, Dario",
    title = "{The general relativistic geometry of the Navarro - Frenk - White model}",
    eprint = "astro-ph/0303594",
    archivePrefix = "arXiv",
    reportNumber = "CIEA-GR-03-35",
    journal = "Rev. Mex. Fis.",
    volume = "51",
    pages = "71--75",
    year = "2005"
}

@article{Vagnozzi:2022moj,
    author = "Vagnozzi, Sunny and others",
    title = "{Horizon-scale tests of gravity theories and fundamental physics from the Event Horizon Telescope image of Sagittarius A}",
    eprint = "2205.07787",
    archivePrefix = "arXiv",
    primaryClass = "gr-qc",
    reportNumber = "UCI-HEP-TR-2022-07",
    doi = "10.1088/1361-6382/acd97b",
    journal = "Class. Quant. Grav.",
    volume = "40",
    number = "16",
    pages = "165007",
    year = "2023"
}

@article{Johnson:2023ynn,
    author = "Johnson, Michael D. and others",
    title = "{Key Science Goals for the Next-Generation Event Horizon Telescope}",
    eprint = "2304.11188",
    archivePrefix = "arXiv",
    primaryClass = "astro-ph.HE",
    doi = "10.3390/galaxies11030061",
    journal = "Galaxies",
    volume = "11",
    number = "3",
    pages = "61",
    year = "2023"
}

@article{Doeleman:2023kzg,
    author = "Doeleman, Sheperd S. and others",
    title = "{Reference Array and Design Consideration for the Next-Generation Event Horizon Telescope}",
    eprint = "2306.08787",
    archivePrefix = "arXiv",
    primaryClass = "astro-ph.IM",
    doi = "10.3390/galaxies11050107",
    journal = "Galaxies",
    volume = "11",
    number = "5",
    pages = "107",
    year = "2023"
}

@article{Kimura:2021dsa,
    author = "Kimura, Masashi and Harada, Tomohiro and Naruko, Atsushi and Toma, Kenji",
    title = "{Backreaction of mass and angular momentum accretion on black holes: General formulation of metric perturbations and application to the Blandford{\textendash}Znajek process}",
    eprint = "2105.05581",
    archivePrefix = "arXiv",
    primaryClass = "gr-qc",
    reportNumber = "RUP-21-7, YITP-21-43",
    doi = "10.1093/ptep/ptab101",
    journal = "PTEP",
    volume = "2021",
    number = "9",
    pages = "093E03",
    year = "2021"
}

@article{Dyson:2025dlj,
    author = "Dyson, Conor and Spieksma, Thomas F. M. and Brito, Richard and van de Meent, Maarten and Dolan, Sam",
    title = "{Environmental Effects in Extreme-Mass-Ratio Inspirals: Perturbations to the Environment in Kerr Spacetimes}",
    eprint = "2501.09806",
    archivePrefix = "arXiv",
    primaryClass = "gr-qc",
    doi = "10.1103/PhysRevLett.134.211403",
    journal = "Phys. Rev. Lett.",
    volume = "134",
    number = "21",
    pages = "211403",
    year = "2025"
}

@article{LaHaye:2025ley,
    author = "LaHaye, Michael and Weller, Colin and Li, Dongjun and Bourg, Patrick and Chen, Yanbei and Yang, Huan",
    title = "{Evolving extreme mass-ratio inspirals in a perturbed Schwarzschild spacetime}",
    eprint = "2510.16102",
    archivePrefix = "arXiv",
    primaryClass = "gr-qc",
    doi = "10.1103/dy1z-36w5",
    journal = "Phys. Rev. D",
    volume = "113",
    number = "2",
    pages = "024069",
    year = "2026"
}

@article{Gliorio:2025cbh,
    author = "Gliorio, Sara and Berti, Emanuele and Maselli, Andrea and Speeney, Nicholas",
    title = "{Extreme mass ratio inspirals in dark matter halos: Dynamics and distinguishability of halo models}",
    eprint = "2503.16649",
    archivePrefix = "arXiv",
    primaryClass = "gr-qc",
    doi = "10.1103/dw6c-14pt",
    journal = "Phys. Rev. D",
    volume = "112",
    number = "12",
    pages = "124050",
    year = "2025"
}

@article{Xavier:2023exm,
    author = "Xavier, S{\'e}rgio V. M. C. B. and Lima, Junior., Haroldo C. D. and Crispino, Lu{\'\i}s C. B.",
    title = "{Shadows of black holes with dark matter halo}",
    eprint = "2303.17666",
    archivePrefix = "arXiv",
    primaryClass = "gr-qc",
    doi = "10.1103/PhysRevD.107.064040",
    journal = "Phys. Rev. D",
    volume = "107",
    number = "6",
    pages = "064040",
    year = "2023"
}

@article{Yang:2023tip,
    author = {Yang, Yi and Liu, Dong and {\"O}vg{\"u}n, Ali and Lambiase, Gaetano and Long, Zheng-Wen},
    title = "{Black hole surrounded by the pseudo-isothermal dark matter halo}",
    eprint = "2308.05544",
    archivePrefix = "arXiv",
    primaryClass = "gr-qc",
    doi = "10.1140/epjc/s10052-024-12412-6",
    journal = "Eur. Phys. J. C",
    volume = "84",
    number = "1",
    pages = "63",
    year = "2024"
}

@article{Ma:2022jsy,
    author = "Ma, Shi-Jie and Ma, Tian-Chi and Deng, Jian-Bo and Hu, Xian-Ru",
    title = "{Shadow of Schwarzschild black hole in the cold dark matter halo}",
    eprint = "2206.12820",
    archivePrefix = "arXiv",
    primaryClass = "gr-qc",
    doi = "10.1142/S0217732323501043",
    journal = "Mod. Phys. Lett. A",
    volume = "38",
    number = "24n25",
    pages = "2350104",
    year = "2023"
}

@article{Capozziello:2023tbo,
    author = "Capozziello, Salvatore and Zare, Soroush and Nieto, Luis M. and Hassanabadi, Hassan",
    title = "{Modified Kerr black holes surrounded by dark matter spike}",
    eprint = "2311.12896",
    archivePrefix = "arXiv",
    primaryClass = "gr-qc",
    doi = "10.1016/j.dark.2025.102065",
    journal = "Phys. Dark Univ.",
    volume = "50",
    pages = "102065",
    year = "2025"
}

@article{Fonseca:2025ehf,
    author = "Fonseca, Dylan S. and Macedo, Caio F. B. and Malato Corr{\^e}a, Mateus and Rubiera-Garcia, Diego",
    title = "{Matter environments around black holes: geodesics, light rings and ultracompact configurations}",
    eprint = "2512.22267",
    archivePrefix = "arXiv",
    primaryClass = "gr-qc",
    month = "12",
    year = "2025"
}

@article{Vertogradov:2024qpf,
    author = {Vertogradov, Vitalii and {\"O}vg{\"u}n, Ali},
    title = "{Analyzing the influence of geometrical deformation on photon sphere and shadow radius: A new analytical approach {\textemdash} Spherically symmetric spacetimes}",
    eprint = "2404.04046",
    archivePrefix = "arXiv",
    primaryClass = "gr-qc",
    doi = "10.1016/j.dark.2024.101541",
    journal = "Phys. Dark Univ.",
    volume = "45",
    pages = "101541",
    year = "2024"
}

@article{Pantig:2024rmr,
    author = "Pantig, Reggie C.",
    title = "{Apparent and emergent dark matter around a Schwarzschild black hole}",
    eprint = "2405.07531",
    archivePrefix = "arXiv",
    primaryClass = "gr-qc",
    doi = "10.1016/j.dark.2024.101550",
    journal = "Phys. Dark Univ.",
    volume = "45",
    pages = "101550",
    year = "2024"
}

@article{Donmez:2024luc,
    author = "Donmez, Orhan and Dogan, Fatih",
    title = "{Estimating the possible QPOs of M87{\ensuremath{*}} from the parameters of a hairy Kerr black hole}",
    eprint = "2407.01478",
    archivePrefix = "arXiv",
    primaryClass = "gr-qc",
    doi = "10.1016/j.dark.2024.101718",
    journal = "Phys. Dark Univ.",
    volume = "46",
    pages = "101718",
    year = "2024"
}

@article{Ovgun:2025bol,
    author = {{\"O}vg{\"u}n, Ali and Pantig, Reggie C.},
    title = "{Black hole in the Dekel-Zhao dark matter profile}",
    eprint = "2501.12559",
    archivePrefix = "arXiv",
    primaryClass = "gr-qc",
    doi = "10.1016/j.physletb.2025.139398",
    journal = "Phys. Lett. B",
    volume = "864",
    pages = "139398",
    year = "2025"
}

@article{Lobo:2025kzb,
    author = "Lobo, Francisco S. N. and Ramos, Jorde A. A. and Rodrigues, Manuel E.",
    title = "{Supermassive black hole in NGC 4649 (M60) with a dark matter halo: impact on shadow measurements and thermodynamic properties}",
    eprint = "2505.03661",
    archivePrefix = "arXiv",
    primaryClass = "gr-qc",
    doi = "10.1088/1475-7516/2025/09/024",
    journal = "JCAP",
    volume = "09",
    pages = "024",
    year = "2025"
}

\end{document}